%% file: nv_paper.tex
\def\CommentVersion{}
\def\ReportVersion{}

\documentclass[10pt,conference]{IEEEtran}
\ifCLASSOPTIONcompsoc
  \usepackage[nocompress]{cite}
\else
  \usepackage{cite}
\fi

\usepackage{amsthm}

\usepackage{amsmath}
\usepackage{graphicx}
\usepackage{tikz}
\usetikzlibrary{arrows,fit,shapes}
\usetikzlibrary{calc}
\tikzset{fontscale/.style = {font=\relsize{#1}}}
\usepackage{pgfplots}
\usepackage{algpseudocode}
\usepackage{algorithm}
\usepackage{caption}
\usepackage{multirow}
\usepackage{subcaption}
\usepackage{relsize}
\usepackage{url}
\usepackage{adjustbox}

\usepackage{pgfplots}
\usepackage{filecontents}
\pgfplotsset{compat=newest}

\usepackage{cite}

\ifdefined\CommentVersion%
  \usepackage{marginnote}
  \usepackage{color}
  \usepackage{soul}
  \newcommand{\marginX}{\marginnote{\huge{\quad\quad\textbf{!}\quad\quad}}}
  \newcommand{\he}[1]{{\color{green}\marginX{}[\textbf{Houssam}: #1]}}
  \newcommand{\gl}[1]{{\color{magenta}\marginX{}[\textbf{Giuseppe}: #1]}}

  \setstcolor{blue}

  \newcommand{\deleted}[1]{\st{#1}}
\else
  \newcommand{\he}[1]{}
  \newcommand{\gl}[1]{}
  \newcommand{\ro}[1]{}
  
  \newcommand{\deleted}[1]{}
  
\fi

\ifdefined\ReportVersion%
  \newcommand{\paperOnly}[1]{}
  \newcommand{\reportOnly}[1]{#1}
\else
  \newcommand{\paperOnly}[1]{#1}
  \newcommand{\reportOnly}[1]{}
\fi

\newtheorem{example}{Example}
\newtheorem{definition}{Definition}
\newtheorem{theorem}{Theorem}
\newtheorem{lemma}{Lemma}

\usepackage{proof}
\usepackage{capt-of}
\usepackage{mathtools}
\usepackage{amsmath,amsfonts,amssymb}
\usepackage{microtype}
\usepackage{hyperref}
\usetikzlibrary{plotmarks}

\makeatletter
\tikzset{
    scale plot marks/.is choice,
    scale plot marks/false/.code={
        \def\pgfuseplotmark##1{\pgftransformresetnontranslations\csname pgf@plot@mark@##1\endcsname}
    },
    scale plot marks/true/.style={},
    scale plot marks/.default=true
}

\makeatother

\newcommand{\task}[1]{\tau_{#1}}
\newcommand{\taskset}{\mathcal{T}}
\newcommand{\vertex}[2]{v_{#1#2}}
\newcommand{\vertexf}{v}
\newcommand{\vertexset}[1]{\mathcal{V}_{#1}}
\newcommand{\period}[1]{\mathsf{T}({#1})}

\newcommand{\offset}[1]{\mathsf{O}({#1})}
\newcommand{\ldeadline}[1]{\mathsf{D}({#1})}

\newcommand{\edgeset}[1]{\mathcal{E}_{#1}}

\newcommand{\charge}[1]{\mathsf{C}(#1)}
\newcommand{\concrete}[1]{\overline{#1}}

\begin{document}

\title{A C-DAG task model for scheduling complex real-time tasks on heterogeneous platforms: preemption matters}


\author{Houssam-Eddine~Zahaf, Nicola~Capodieci, Roberto~Cavicchioli, Marko~Bertogna, Giuseppe~Lipari }



\IEEEtitleabstractindextext{
  \input{texfiles/abstract.tex}
}

\maketitle

\IEEEdisplaynontitleabstractindextext

\IEEEpeerreviewmaketitle

\input{texfiles/intro.tex}
\input{texfiles/models.tex}
\input{texfiles/analysis.tex}

\input{texfiles/allocation.tex}
\input{texfiles/stat.tex}
\input{texfiles/results.tex}

\input{texfiles/conclu.tex}

\bibliographystyle{IEEEtran}
\bibliography{mybib}

\end{document}

%% file: texfiles/abstract.tex
\begin{abstract}
  \label{sec:abstract} 
  Recent commercial hardware platforms for embedded real-time systems
  feature heterogeneous processing units and computing accelerators on
  the same System-on-Chip. 
  When designing complex real-time application for such architectures, 
  the designer needs to make a number of difficult choices: on which
  processor should a certain task be implemented? Should a component
  be implemented in parallel or sequentially? These choices may have a
  great impact on feasibility, as the difference in the processor
  internal architectures impact on the tasks' execution time and
  preemption cost.

  To help the designer explore the wide space of design choices and
  tune the scheduling parameters, in this paper we propose a novel
  real-time application model, called C-DAG, specifically conceived
  for heterogeneous platforms. A C-DAG allows to specify alternative
  implementations of the same component of an application for
  different processing engines to be selected off-line, as well as
  conditional branches to model if-then-else statements to be selected
  at run-time.

  We also propose a schedulability analysis for the C-DAG model and a
  heuristic allocation algorithm so that all deadlines are
  respected. Our analysis takes into account the cost of preempting a
  task, which can be non-negligible on certain processors. We
  demonstrate the effectiveness of our approach on a large set of
  synthetic experiments by comparing with state of the art algorithms
  in the literature.
\end{abstract}


%% file: texfiles/intro.tex
\section{Introduction} 
\label{sec:introduction}

Modern cyber-physical embedded systems demand are increasingly complex
and demand powerful computational hardware platforms. A recent trend
in hardware architecture design is to combine high performance
multi-core CPU hosts with a number of application-specific
accelerators (e.g. Graphic Processing Units -- GPUs, Deep Learning
Accelerators -- DLAs, or FPGAs for programmable hardware) in order to
support complex real-time applications with machine learning and image
processing software modules.

Such application specific processors are defined by different levels
of programmability and a different Instruction Set Architecture (ISA)
compared to the more traditional SoCs.  NVIDIA Volta GPU architecture
for instance\footnote{NVIDIA GV100 White Paper
  \url{http://images.nvidia.com/content/volta-architecture/pdf/volta-architecture-whitepaper.pdf}},
couples a fairly traditional GPU architecture (hundreds of small SIMD
processing units called \emph{CUDA cores}, grouped in computing
clusters called \emph{Streaming Multiprocessors}) with hardware
pipelines specifically designed for tensor processing (\emph{Tensor
  Cores}), hence designed for \emph{matrix multiply and accumulate}
operations that are typical of neural network arithmetics. The
integrated version of the NVIDIA Volta architecture is embedded within
the NVIDIA Xavier SoC, which can now be found in the NVIDIA Jetson AGX
board and in the NVIDIA Pegasus board: in such embedded platforms,
tensor processing can also be operated in specifically designed
compute engines such as the DLA (Deep Learning
Accelerator\footnote{Hardware specifications for the DLA available at
  \url{http://nvdla.org/}}); moreover, another application specific
engine is the PVA (Programmable Vision Accelerator), that is
specifically designed for solving signal processing algorithms such as
stereo disparity and optical flow.

In such platforms, the main and novel challenge in analyzing the
timing behavior of a real-time application is represented by the
drastic differences at the level of ISAs, preemption capabilities,
memory hierarchies and inter-connections for these collections of
computing engines.

When programming these platforms, the software designer is confronted
with several design choices: on which processor engine should a task
be implemented? Should a certain sub-system be implemented in parallel
or sequentially? These choices could impact on the timing behavior of
the application and on the resource utilization. The analysis is
complicated by the fact that, on certain processors, the overhead
induced by preempting a lower priority task can be large: for example,
the overhead of preempting a graphical task executing on certain GPU
architectures is in the same order of magnitude of the worst-case
execution time of the task. As we will see in Section
\ref{sec:preempt-cost-simul}, such overhead depends on the computing
engine and on the type of task.

\paragraph{Contributions.}
To help the designer explore the design space, in
Section~\ref{sec:system-model} we present a novel model of real-time
task called C-DAG (Conditional-Directed Acyclic Graph). Thanks to the
graph structure, the C-DAG model allows to specify parallelism of
real-time sub-tasks. The designer can use special \emph{alternative}
nodes in the graph to model alternative implementations of the same
functionality on different computing engines to be selected off-line,
and \emph{conditional} nodes in the graph to model if-then-else
branches to be selected at run-time. Alternative nodes are used to
leverage the diversity of computing accelerators within our target
platform.

Then, in Section \ref{sec:analysis} we present a schedulability
analysis that will be used in Section \ref{sec:allocation} by a set of
allocation heuristics to map tasks on computing platforms and to
assign scheduling parameters. In particular, we present a novel
technique to reduce the pessimism due to high preemption costs in the
analysis (Section~\ref{sec:preempt-aware-sched}).

After discussing related work in Section \ref{sec:related-work}, our
methodology is evaluated in Section~\ref{sec:results} by comparing it
with start of the art algorithms trough a set of synthetic
experiments.








%% file: texfiles/models.tex
\section{System model}
\label{sec:system-model}

\subsection{Architecture model}
\label{sec:architecture-model}

A heterogeneous architecture is modeled as a set of \emph{execution
  engines} $\mathtt{Arch} = \{ e_{1}, e_2, \ldots, e_m\}$.  An
execution engine is characterized by 1) its execution capabilities,
(i.e. its Instruction Set Architecture), specified by the engine's
\emph{tag}, and 2) its scheduling policy. An engine's tag
$\mathtt{tag}(e_i)$ indicates the ability of a processor to execute a
dedicated tasks.

As an example, a Xavier based platform such as the \emph{NVIDIA
  pegasus}, can be modeled using a total of $16$ engines for a total
of five different engine tags: $8$ \textsf{CPU}s, $2$ \textsf{dGPU}s,
$2$ \textsf{iGPU}s, $2$ \textsf{DLA}s and $2$ \textsf{PVA}s.

Tags express the heterogeneity of modern processor architecture: an
engine tagged by \textsf{dGPU} (discrete GPU) or \textsf{iGPU}
(integrated GPU) is designed to efficiently run generic GPU kernels,
whereas engines with \textsf{DLA} tags are designed to run \emph{deep
  learning inference} tasks.

Trivially, a deep learning task can be compiled to run on any engine,
including CPUs and GPUs, however its worst-case execution time will be
lower when running on DLAs. In this paper, we allow the designer to
compile the same task on different alternative engines with different
tradeoffs in terms of performance and resource utilization, so to
widen the space of possible solutions. As we will see in the next
section, the C-DAG model supports \emph{alternative} implementations
of the same code. During the off-line analysis phase, only one of
these alternative versions will be chosen depending on the overall
schedulability of the system.



Engines are further characterized by a scheduling policy (e.g. Fixed
Priority or Earliest Deadline First), which can be \emph{preemptive}
or \emph{non-preemptive}. In our model we allow different engines to
support different scheduling policies: as we show in
Section~\ref{sec:analysis}, in our methodology the schedulability
analysis of each engine can be performed independently of the
others. However, to simplify the presentation, in this paper we focus
only on \emph{preemptive EDF} for all the considered engines.

\subsection{The C-DAG task model}
\label{sec:task-model}

\subsubsection{Specification tasks}

A \emph{specification task} is a Directed Acyclic Graph (DAG),
characterized by a tuple
$\tau = \{\mathsf{T}, \mathsf{D}, \vertexset{}, \mathcal{A}, \Gamma,
\edgeset{}\}$,
where: $\mathsf{T}$ is the period (minimum interarrival time);
$\mathsf{D}$ is the relative deadline; $\vertexset{}$ is a set of
graph nodes that represent \emph{sub-tasks}; $\mathcal{A}$ is a set of
\emph{alternative nodes}; and $\Gamma$ is a set of \emph{conditional
  nodes}. The set of all the nodes is denoted by
$\mathcal{N} = \vertexset{} \cup \mathcal{A} \cup \Gamma$. The set
$\edgeset{}$ is the set of edges of the graph
$\edgeset{} : \mathcal{N} \times \mathcal{N}$.

A sub-task $v \in \vertexset{}$ is the basic computation unit.
It represents a block of code to be executed by one of the engines of
the architecture. A sub-task is characterized by:
\begin{itemize}
\item A tag $\mathtt{tag}(v)$ represent the ISA of the sub-task
  code. A sub-task can only be allocate onto an engine with the same
  tag;
\item A worst-case execution time $C(v)$ when executing the sub-task
  on the corresponding engine processor.
\end{itemize}

A conditional node $\gamma \in \Gamma$ represents alternative paths in
the graph due to non-deterministic on-line conditions
(e.g. if-then-else conditions). At run-time, only one of the outgoing
edges of $\gamma$ is executed, but it is not possible to know in
advance which one.

An alternative node $a \in \mathcal{A}$ represents alternative
implementations of parts of the graph/task, as introduced in the
previous section. During the configuration phase (which is detailed in
Section~\ref{sec:vertex-allocation}) our methodology selects one
between many possible alternative implementations of the program by
selecting only one of the outgoing edges of $a$ and removing (part of)
the paths starting from the other edges. This can be useful when
modeling sub-tasks than can be executed on different engines with
different execution costs.  In our model, the choice of where the
sub-task should be executed is performed off-line by our proposed
scheduling analysis and allocation strategy.

An edge $e(n_i, n_j) \in \edgeset{}$ models a precedence constraint
(and related communication) between node $n_i$ and node $n_j$, where
$n_i$ and $n_j$ can be sub-tasks, alternative nodes or conditional
nodes. 

The set of \emph{immediate predecessors} of a node $n_j$, denoted by
$\mathsf{pred}(n_j)$, is the set of all nodes $n_i$ such that there
exists an edge $(n_i, n_j)$. The set of \emph{predecessors} of a node
$n_j$ is the set of all nodes for which there exist a path toward
$n_j$.  If a node has no predecessor, it is a \emph{source node} of
the graph. In our model we allow a graph to have several source nodes.
In the same way we can define the set of \emph{immediate successors}
of node $n_j$, denoted by $\mathsf{succ}(n_j)$, as the set of all
nodes $n_k$ such that there exists an edge $(n_j, n_k)$, and the set
of \emph{successors} of $n_j$ as the set of nodes for which there is
a path from $n_j$. If a node has no successors, it is a \emph{sink
  node} of the graph, and we allow a graph to have several sink nodes.

Conditional nodes and alternative nodes always have at least 2
outgoing edges, so they cannot be sinks. To simplify the reasoning, we
also assume that they always have at least one predecessor node, so
they cannot be sources.

\subsubsection{Concrete tasks}

A concrete task
$\concrete{\tau} = \{T, D, \concrete{\vertexset{}}, \concrete{\Gamma},
\concrete{\edgeset{}} \}$ is an instance of a specification task where
all alternatives have been removed by making implementation
choices 
during the analysis.
Before explaining how to obtain a concrete task from a specification
task, we present an example.

\begin{example}
\label{example:task-model-1}
Consider the task specification described in 
Figure~\ref{fig:tau_spec_2}. Each sub-task node is labeled by the
sub-task id and engine tag. 
Alternative nodes are denoted by square boxes and conditional nodes
are denoted by diamond boxes. The black boxes denote corresponding
junction nodes for alternatives and conditional, they are used to
improve the readability of the figure but they are not part of the
task specification\footnote{In fact, it is not always possible to insert
junction nodes for an arbitrary specification.}.

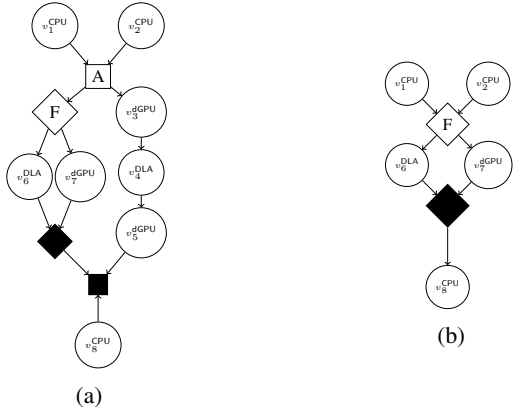
\begin{figure}[h]
  \centering
  \begin{subfigure}{.5\columnwidth}
    \centering
    \resizebox{0.5\textwidth}{!}{\input{figs/task_spec.tex}}
    \caption{}\label{fig:tau_spec_2}
  \end{subfigure}
  \quad\quad        
  \begin{subfigure}{.4\columnwidth}
    \centering
    \resizebox{0.5\textwidth}{!}{\input{figs/concrete_1.tex}}
    \caption{}\label{fig:concrete_1}
  \end{subfigure}
  \caption{Task specification and concrete tasks}
\end{figure}


Sub-tasks $\vertex{}{1}^{\mathsf{CPU}}$ and
$\vertex{}{2}^{\mathsf{CPU}}$ are the sources (entry points) of the
DAG. $\vertex{}{1}^{\mathsf{CPU}}, \vertex{}{2}^{\mathsf{CPU}}$ are
marked by the \textsf{CPU} tag and can run cuncurrently: during the
off-line analysis they may be allocated on the same or onto different
engines. Sub-task $\vertex{}{4}^{\mathsf{DLA}}$ has an outgoing edge
to $\vertex{}{5}^{\mathsf{dGPU}}$, thus sub-task
$\vertex{}{5}^{\mathsf{dCPU}}$ can not start its execution before
sub-task $\vertex{}{4}^{\mathsf{DLA}}$ has finished its
execution. Sub-tasks $\vertex{}{1}^{\mathsf{CPU}}$ and
$\vertex{}{2}^{\mathsf{CPU}}$ have each one outgoing edge to the
alternative node $A$. Thus, $\task{}$ can execute either:
\begin{enumerate}
\item by following $\vertex{}{3}^{\mathsf{dGPU}}$ and then
  $\vertex{}{4}^{\mathsf{DLA}}$,$\vertex{}{5}^{\mathsf{dGPU}}$ and
  finishing its instance on $\vertex{}{8}^{\mathsf{CPU}}$;
\item or by following the conditional node $F$ and select, according
  to an undetermined condition evaluated on-line, either to execute
  $\vertex{}{6}^{\mathsf{DLA}}$ or $\vertex{}{7}^{\mathsf{dGPU}}$, and
  finishing its instance on $\vertex{}{8}^{\mathsf{CPU}}$.
\end{enumerate}

The two patterns are alternative ways to execute the same
functionalities at different costs. 

Figure~\ref{fig:concrete_1} represents one of the concrete tasks of
$\task{i}$.  During the analysis, alternative execution
$(\vertex{}{3}^{dGPU},\vertex{}{4}^{DLA},\vertex{}{5}^{dGPU})$ has
been dropped.
\end{example}

We consider a \emph{sporadic task} model, therefore parameter
$\mathsf{T}$ represents the minimum inter-arrival times between two
instances of the same concrete task. When an instance of a task is
activated at time $t$, all source sub-tasks are simultaneously
activated. All subsequent sub-tasks are activated upon completion of
their predecessors, and sink sub-tasks must all complete no later
than time $t+\mathsf{D}$. We assume \emph{constrained deadline tasks},
that is $\mathsf{D} \leq \mathsf{T}$.

We now present a procedure to generate a concrete task
$\concrete{\tau}$ from a specification task $\tau$, when all
alternatives have been chosen. The procedure starts by initializing
$\concrete{\vertexset{}} = \emptyset$,
$\concrete{\Gamma} = \emptyset$. First, all the source sub-tasks of
$\tau$ are added to $\concrete{\vertexset{}}$.  Then, for every
immediate successor node $n_j$ of a node
$n_i \in \{\concrete{\vertexset{}} \cup \concrete{\Gamma} \}$: if
$n_j$ is a sub-task node (a conditional node), it is added to
$\concrete{\vertexset{}}$ (to $\concrete{\Gamma}$, respectively); if
it is an alternative node, we consider the selected immediate
successor $n_k$ of $n_j$ and we add it to $\concrete{\vertexset{}}$ or
to $\concrete{\Gamma}$, respectively. The procedure is iterated until
all nodes of $\tau$ have been visited. The set of edges
$\concrete{\edgeset{}} \subseteq \edgeset{}$ is updated accordingly.

We denote by $\Omega(\tau)$ the set of all concrete tasks of a
specification task $\tau$. $\Omega(\tau)$ is generated by simply
enumerating all possible alternatives. 


%% file: figs/task_spec.tex
\begin{tikzpicture}
  \def\d{1.7}
  \node at (0,0) [circle,draw] (v_1) {\tiny $v_{1}^{\mathsf{CPU}}$};
  \node at (\d,0) [circle, draw] (v_2) {\tiny $v_{2}^{\mathsf{CPU}}$};
  \node at (\d/2,-1) [rectangle,draw] (F_1) {\textcolor{black}{A}};
  \node at (\d/2,-3*\d) [rectangle,draw,fill] (F_4) {\textcolor{black}{\scriptsize F}};
  \node at (\d,-\d) [circle, draw] (v_4) {\tiny $v_{3}^{\mathsf{dGPU}}$};
  \node at (\d,-1.7*\d) [circle, draw] (v_5) {\tiny $v_{4}^{\mathsf{DLA}}$};
  \node at (\d,-2.4*\d) [circle, draw] (v_6) {\tiny $v_{5}^{\mathsf{dGPU}}$};
  \node at (\d/2,-3.7*\d) [circle, draw] (v_10) {\tiny $v_{8}^{\mathsf{CPU}}$};
  \node at (0,-\d) [diamond,draw] (F_2) {F};
  \node at (0,-2.5*\d) [diamond,draw,fill] (F_3) {\tiny F};
  \node at (-0.5,-1.75*\d) [circle, draw] (v_7) {\tiny $v_{6}^{\mathsf{DLA}}$};
  \node at (0.5,-1.75*\d) [circle, draw] (v_8) {\tiny $v_{7}^{\mathsf{dGPU}}$};
  \draw [<-](F_1) -- (v_1);
  \draw [<-](F_1) -- (v_2);
  \draw [<-](F_2) -- (F_1);
  \draw [->](F_1) -- (v_4);
  \draw [<-](v_5) -- (v_4);
  \draw [->](v_5) -- (v_6);
  \draw [->](F_2) -- (v_7);
  \draw [->](F_2) -- (v_8);
  \draw [<-](F_3) -- (v_7);
  \draw [<-](F_3) -- (v_8);
  \draw [->](F_3) -- (F_4);
  \draw [->](v_6) -- (F_4);
  \draw [->](v_10) -- (F_4);
\end{tikzpicture}

%% file: figs/concrete_1.tex
\begin{tikzpicture}
  \def\d{1.7}
  \node at (0,0) [circle,draw] (v_1) {\tiny $v_{1}^{\mathsf{CPU}}$};
  \node at (\d,0) [circle, draw] (v_2) {\tiny $v_{2}^{\mathsf{CPU}}$};
  \node at (\d/2,-0.5*\d) [diamond,draw] (F_2) {F};
  \node at (\d/2,-2.5*\d) [circle, draw] (v_10) {\tiny $v_{8}^{\mathsf{CPU}}$};
  
  \node at (\d/2,-1.5*\d) [diamond,draw,fill] (F_3) {F};
  \node at (0,-1*\d) [circle, draw] (v_7) {\tiny $v_{6}^{\mathsf{DLA}}$};
  \node at (\d,-1*\d) [circle, draw] (v_8) {\tiny $v_{7}^{\mathsf{dGPU}}$};
  \draw [<-](F_2) -- (v_1);
  \draw [<-](F_2) -- (v_2);
  \draw [->](F_2) -- (v_7);
  \draw [->](F_2) -- (v_8);
  \draw [<-](F_3) -- (v_7);
  \draw [<-](F_3) -- (v_8);
   \draw [<-](v_10) -- (F_3);
\end{tikzpicture}

%% file: texfiles/analysis.tex
\section{Scheduling analysis}
\label{sec:analysis}

In this work, we consider partitioned scheduling. Each engine has its
own scheduler and a separate ready-queue. Sub-tasks are allocated
(partitioned) onto the available engines so that the system is
schedulable. Partitioned scheduling allows to use well-known single
processor schedulability tests which make the analysis simpler and
allow us to reduce the overhead due to thread migration compared to
global scheduling. The analysis presented here is modular, so engines
may have different scheduling policies. In this paper, we restrict to
preemptive-EDF.

\subsection{Alternative patterns}
\label{sec:graph-spec-transf}

Given a specification task $\tau$, we have to select one of the
possible concrete tasks before proceeding to the allocation and
scheduling of the sub-tasks on the computing engine. Since the number
of combinations can be very large, in this paper we propose an
heuristic algorithm based on a \emph{greedy} strategy (see
Section~\ref{sec:allocation}). In particular, we explore the set of
concrete tasks in a certain order. The order relation $\succ$ sorts
concrete tasks according to their total execution time.

\begin{definition}
  Let $\concrete{\task{}}',\concrete{\task{}}''$ be two concrete tasks
  of specification task $\task{}$

  The partial order relation $\succ$ is defined as:

  \begin{equation}
    \label{eq:ord_1}
    \concrete{\task{}}' \succ \concrete{\task{}}'' \implies C(\concrete{\tau}') \geq C(\concrete{\tau}'')
  \end{equation}
\end{definition}

In the next section, we will define a second order relationship $\gg$
that sorts concrete tasks based on their engine tags.

\subsection{Tagged Tasks}
\label{sec:graph-separation}

One concrete task may contain sub-tasks with different tags which will
be allocated on different engines. 
Before proceeding to allocation, we need to select only sub-tasks
pertaining to a given tag. We call this operation \emph{task
  filtering}.

We start by defining an \emph{empty sub-task} as a sub-task with null
computation time.
\begin{definition}[Tagged task]\label{def:tagged-task}
  Let
  $\concrete{\tau} = \{\mathsf{T}, \mathsf{D},
  \concrete{\vertexset{}}, \concrete{\Gamma}, \concrete{\edgeset{}}\}$
  be a concrete task. Task $\concrete{\tau}(\mathsf{tag}_i)$ is a
  \emph{tagged task} of $\concrete{\tau}$ iff 
  \begin{itemize}
  \item $\concrete{\tau}(\mathsf{tag}_i) = \{\mathsf{T}, \mathsf{D},
    \vertexset{i}, \Gamma_i, \edgeset{i}\}$
    is isomorphic to $\concrete{\tau}$, that is the graph has the same
    structure, the same number of nodes of the same type, and the same
    edges between corresponding nodes;


  \item let $v \in \concrete{\vertexset{}}$ be a sub-task of
    $\concrete{\tau}$, and let $v' \in \vertexset{i}$ be the
    corresponding sub-task of $\concrete{\tau}(\mathsf{tag}_i)$ in the
    isomorphism. If $\mathsf{tag}(v) = \mathsf{tag}_i$, then
    $C(v') = C(v)$, else $C(v') = 0$;

  \item $\Gamma_i = \concrete{\Gamma}$.
  \end{itemize}
  We denote with
  $\mathcal{S}(\concrete{\tau}) = \{ \concrete{\tau}(\mathsf{tag}_1),
  \ldots \concrete{\tau}(\mathsf{tag}_K) \}$
  the set of all possible tagged tasks of $\concrete{\tau}$.
\end{definition}

Each concrete task generates as many \emph{tagged tasks} as there are
tags in the target architecture. 


\begin{figure}[h]
  \centering
  \begin{subfigure}{.29\columnwidth}
    \centering
    \resizebox{0.95\textwidth}{!}{\input{figs/sep_1}}
  \end{subfigure}
\quad
  \begin{subfigure}{.29\columnwidth}
    \centering
    \resizebox{0.95\textwidth}{!}{\input{figs/sep_2}}
  \end{subfigure}
  \quad
    \begin{subfigure}{.29\columnwidth}
    \centering
    \resizebox{0.95\textwidth}{!}{\input{figs/sep_3}}
  \end{subfigure}

  \caption{Tagged tasks for the concrete task of
    Figure \label{fig:sep_example}}
\end{figure}
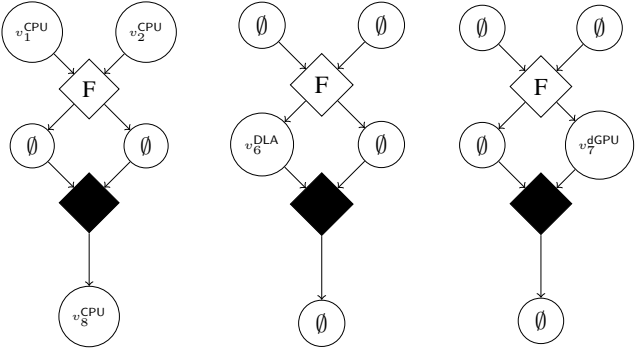

Figure \ref{fig:sep_example} shows the three tagged tasks for the
concrete task in Figure \ref{fig:concrete_1}. 
The first one contains only sub-tasks having CPU tag, the second
contains only DLA sub-tasks, and the third one refers to GPU
sub-tasks. Every tagged task will be allocated on one or more engines
having the corresponding tag.

\begin{definition}[$\gg$ order relationship]
  Assume the architecture supports $K$ different tags.  Let
  $n(\mathsf{tag})$ denote the number of computing engines labeled
  with $\mathsf{tag}$. Assume that tags are ordered by increasing
  $n(\mathsf{tag})$, that is
  $n(\mathsf{tag}_i) < n(\mathsf{tag}_j) \implies i < j$.

  Let $\concrete{\task{}}', \concrete{\task{}}''$ be two concrete
  tasks of specification task $\task{}$, and let
  $\mathcal{S}(\concrete{\tau}') = \{
  \concrete{\tau}'(\mathsf{tag}_1), \ldots,
  \concrete{\tau}'(\mathsf{tag}_K)\}$
  and
  $\mathcal{S}(\concrete{\tau}'') = \{
  \concrete{\tau}''(\mathsf{tag}_1), \ldots,
  \concrete{\tau}''(\mathsf{tag}_K)\}$ be the respective tagged tasks.

  The order relation $\concrete{\task{}}' \gg \concrete{\task{d}}''$
  is defined as follows: 
  \begin{align*}
    & \concrete{\task{}}' \gg \concrete{\task{d}}'' \implies\\
    & \exists \; 0 \leq i \leq K \; \left \{ 
         \begin{aligned}
             & C(\concrete{\tau}'(\mathsf{tag}_j)) = C(\concrete{\tau}''(\mathsf{tag}_j))  \quad \forall j < i \\
             & C(\concrete{\tau}'(\mathsf{tag}_i)) < C(\concrete{\tau}''(\mathsf{tag}_i))
         \end{aligned}\right .
  \end{align*}
\end{definition}

Relationship $\gg$ gives priority to concrete tasks that allocate less
load on scarce resources: if there are few execution engines with
a certain tag, and there is a large number of sub-tasks requiring
allocation on that specific engine, the relation order prefers
alternative patterns with lower workload for those engines. 

\subsection{Deadlines and offsets assignment}
\label{sec:artif-deadl-offs}

Meeting timing constraints of a concrete task depends on the
allocation of the sub-tasks onto the different execution engines. As
these sub-tasks communicate through shared buffers, they are forced to
respect the execution order dictated by the precedence constraints
imposed by the graph structure.

To reduce the complexity of dealing with precedence constraints
directly, we impose intermediate offsets and deadlines on each
sub-task. In this way, precedence constraints are respected
automatically if every sub-task is activated after its offset and it
completes no later than its deadline.

Many authors have proposed techniques to assign intermediate deadlines
and offsets to task graphs. In this paper we use techniques similar to
those proposed in \cite{Marinca:2004:ADA:2285761.2285882} and
\cite{WU2014247}. 

Most of the deadline assignment techniques are based on the
computation of the execution time of the critical path. A path
$P_x = \{v_1,v_2,\cdots, v_l\}$ is a sequence of sub-tasks of task
$\concrete{\task{}}$ such that:
\[ 
  \forall v_l, v_{l+1} \in P_x, \exists e(v_l,v_{l+1}) \in \edgeset{}.
\]
  
Let $\mathcal{P}$ denote the set of all possible paths of task
$\concrete{\tau}$. The critical path
$P_{crit}(\concrete{\tau}) \in \mathcal{P}$ is defined as the path
with the largest cumulative execution time of the sub-tasks.

We define the slack $\mathsf{Sl}(P, \mathsf{D})$ along path $P$ as:
\begin{equation}\nonumber
  \mathsf{Sl}(P, \mathsf{D}) = \mathsf{D} - \sum_{\vertexf \in P} \charge{\vertexf}
\end{equation}

The assignment algorithm starts by assigning an intermediate relative
deadline to every sub-task along a path by distributing the path's
slack as follows:
\begin{equation}\nonumber
  \ldeadline{\vertexf} = \charge{\vertexf} + \mathsf{calculate\_share}(\vertexf, P)
\end{equation}

The \textsf{calculate\_share} function computes the slack for sub-task
$v$ along the path. This slack can be shared according to two
alternative heuristics:
\begin{itemize}
\item \textbf{Fair distribution:} assigns slack as the ratio of the
  original slack by the number of sub-tasks along the path:
  \begin{equation}
    \label{eq:fair-distribution}
   \mathsf{calculate\_share}(\vertexf, P) = \frac{\mathsf{Sl}(P, \mathsf{D})}{|P|}
  \end{equation}
\item \textbf{Proportional distribution:} assigns slack
  according to the contribution of the sub-task execution time in the
  path: 
  \begin{equation}
    \label{eq:prop-distribution}
    \mathsf{calculate\_share}(\vertexf, P) = \frac{\charge{\vertexf}}{\charge{P}} \cdot \mathsf{Sl}(P, \mathsf{D})
  \end{equation}
\end{itemize}

Once the relative deadlines of the sub-tasks along the critical path
have been assigned, we can select the next path in order of decreasing
cumulative execution time, and assign the deadlines to the remaining
sub-task by appropriately subtracting the already assigned
deadlines. The complete procedure has been described in
\cite{WU2014247}, and due to space constraints we do not report it
here.

Let $\offset{\vertexf}$ be the offset of a subtask with respect of the
arrival time of the task's instance. The sum of the offset and of the
intermediate relative deadline of a subtask is called \emph{local
  deadline} $\offset{\vertexf}+\ldeadline{\vertexf}$, and it is the
deadline relative to the arrival of the task's instance.

The offset of a subtask is set equal to 0 if the subtask has no
predecessors; otherwise, it can be computed recursively as the maximum
between the local deadlines of the predecessor sub-tasks.




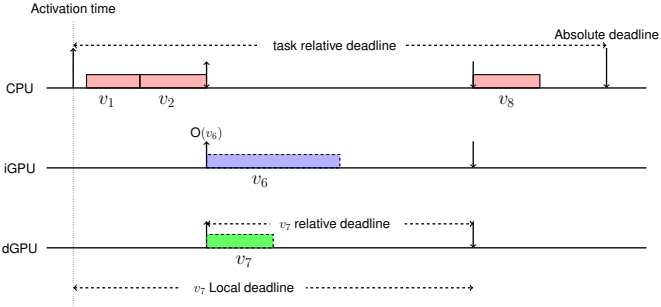
\begin{figure}[h]
  \centering
\resizebox{\columnwidth}{!}{\input{figs/task_off_d.tex}}
  \caption{Example of offset and local deadline}
  \label{fig:task_off_d}
\end{figure}

Figure \ref{fig:task_off_d} illustrates the relationship between the
activation times, the intermediate offsets, relative deadlines and
local deadlines of the sub-tasks of the concrete task of Figure
\ref{fig:concrete_1}. We assume that $v_1,v_2,v_8$ have been allocated
on the same CPU whereas $v_6$ and $v_7$ each on a different
engine. The activation time is the absolute time of the arrival of the
sub-task instance. The activation time of a source sub-task
corresponds to the activation time of the task graph. The offset is
the interval between the activation of the task graph and the
activation of the sub-task. The local deadline is the interval between
the task graph activation and the sub-task absolute deadline.

\begin{definition}
  Sub-task $\vertexf \in \concrete{\vertexset{\tau}}$ is feasible if
  for each task instance arrived at $a_j$, sub-task $\vertexf$
  executes within the interval bounded by its arrival time
  $a(\vertexf) = a_j + \offset{\vertexf}$ and its \emph{absolute}
  deadline $a(\vertexf) + \ldeadline{\vertexf}$.
\end{definition}

\begin{lemma}
  \label{theo:vertices-feasibility}
  A concrete task (resp. tagged task) is feasible if 
  all its sub-tasks are feasible.
\end{lemma}
\begin{proof}
  By the definition, the local deadline of the sink sub-tasks is equal
  to the deadline of the task $\mathsf{D}$. Moreover, the offset of a
  sub-task is never before the local deadline of a preceding
  sub-task. Therefore 1) the precedence constraints are respected and
  2) if sink sub-tasks are feasible then the concrete task (tagged
  task, respectively) is feasible.
\end{proof}

\subsection{Single engine analysis}
\label{sec:single-core-analyis}

In this section, we assume that sub-tasks have been already been
assigned offsets and deadlines, and they have been allocated on the
platform's engines, and we present the schedulability analysis to test
if all tasks respect their deadlines when scheduled by the Earliest
Deadline First (EDF) algorithm.

\begin{theorem}\label{theo:single-engine-analys}
  Let $\taskset{}$ a set of task graphs allocated onto a single-core
  engine. Task set $\taskset{}$ is schedulable by EDF if and only if:
  \begin{equation}\label{eq:dbf-classic}
  \sum_{\concrete{\task{}} \in \taskset} \mathsf{dbf}(\concrete{\task{}}, t) \leq t,  \forall t \leq t^*
  \end{equation}
\end{theorem}
The $\mathsf{dbf}$ is the \textsf{demand bound
  function}~\cite{baruah1990algorithms} for a task graph
$\concrete{\task{}}$ in interval $t$. The demand bound function is
computed as the worst-case cumulative execution time of all jobs
(instances of sub-tasks) having their arrival time and deadline within
any interval of time of length $t$. For a task graph, the
$\mathsf{dbf}$ can be computed as follows:

\begin{equation}
  \label{eq:dbf-elemntary}
  \mathsf{dbf}(\tau, t) = \max_{v \in \tau} \sum_{v' \in \tau} 
  \left\lfloor \frac{t - \tilde{O}(v') - \mathsf{D}(v') + \period{\task{}}}{\period{\task{}}}\right\rfloor \charge{v'}
\end{equation}

\noindent where\footnote{We remind that the remainder of $a / b$ is by
  definition a positive number $r$ such that $a = kb + r$.}:
\begin{align*}\tilde{O}(v') &= (\offset{v'} - \offset{v}) \; \mathsf{mod} \;
\period{\task{}} \\
\end{align*}

In our model, a task graph may contain \emph{conditional nodes}, which
model alternative paths that are selected non-deterministically at
run-time. To compute the \textsf{dbf} for a tagged task that contains
conditional nodes, we must first enumerate all possible conditional
graphs by using the same procedure as the one used for generating
concrete tasks from specification tasks. Hence, the \textsf{dbf} of a
tagged task in interval $t$ can be computed as the largest
$\mathsf{dbf}$ among all the possible conditional graphs.

\subsection{Anticipating the activation of  sub-tasks}
\label{sec:antic-start-time}

Given an instance of sub-task $v$ with arrival at $a(v)$ and local
deadline at $\ldeadline{v}$, at run-time it may happen that all
instances of the preceding sub-tasks have already completed their
execution before $a(v)$. In this case, we activate the sub-task as
soon as the preceding sub-tasks have finished \emph{with the same
  local deadline} $\ldeadline{v}$. 

\begin{lemma}\label{lm:edf-sustainability}
  Consider a feasible set of sub-tasks allocated on a set of engines
  and scheduled by EDF. If a sub-task is activated as soon as all
  predecessor sub-tasks have finished, with the same local deadline,
  the set remains schedulable.
\end{lemma}
\begin{proof}
  Descends directly from the sustainability property of EDF
  \cite{burns2008sustainability}. In fact, by anticipating the
  activation of the sub-task without modifying its local deadline, the
  sub-task will be scheduled with a longer relative deadline, and the
  demand bound function will not increase.
\end{proof}

From an implementation point of view, this technique avoids the need
to set-up activation timers for intermediate tasks; moreover, it
allows us to reduce the pessimism of the analysis in the presence of
high preemption costs, as we will see in the next section.

\subsection{Preemption-aware analysis}
\label{sec:preempt-aware-sched}

In recent GPUs, preempting an executing task can be a costly operation
(see Section \ref{sec:preempt-cost-simul}). In particular, the cost of
preemption may significantly vary depending on the preempted task and
the engine. For example, preempting a graphical kernel induces a
larger cost compared to preempting a computing CUDA kernel. Therefore,
we need to account for the cost of preemption in the analysis.

We start by observing that, in the case of EDF scheduling, a job of a
sub-task $v_i$ can preempt a job of sub-task $v_j$ at most once, and
only if its relative deadline deadline is shorter: $D(v_i) < D(v_j)$.

A simple (although pessimistic) approach is to always consider the
worst-case preemption cost as part of the worst-case computation time
of the preempting task. Let $\mathsf{pc}(v_j)$ denote the cost of
preempting sub-task $v_j$. 

\begin{lemma}\label{theo:worst_preemption}
  Let $\vertexset{}=\{v_{1},v_{2}, \cdots , v_{K}\}$ be a set of
  sub-tasks to be scheduled by EDF on a single engine.

  Consider
  $\vertexset{}^{\mathsf{pc}}=\{v'_{1},v'_{2}, \cdots , v'_{K}\}$,
  where $v'_i$ has the same parameters as $v_i$, except for the wcet
  that is computed as $\charge{v'_i} = \charge{v_i} + \mathsf{pc}^{i}$
  and
  $\mathsf{pc}^{i} = \max\{ \mathsf{pc}(v) | v \in \vertexset{} \wedge
  D(v) > D(v_i) \}$.

  If $\vertexset{}^{\mathsf{pc}}$ is schedulable by EDF when
  considering a null preemption cost, then $\vertexset{}$ is
  schedulable when considering the cost of preemption.
\end{lemma}
\begin{proof}
  The Lemma directly follows from the simple observation that the cost
  of preemption can never exceed $\mathsf{pc}^{i}$ for sub-task $v_i$.
\end{proof}

Lemma~\ref{theo:worst_preemption} is safe but pessimistic. We can
further improve the analysis by observing that a sub-task cannot
preempt another sub-task belonging to the same task graph (we remind
the reader that we assume constrained deadline tasks). Furthermore, it
may be impossible for two consecutive sub-task of a task graph to both
preempt the same sub-task as demonstrated by Theorem
\ref{theo:preempt-all}. 

\begin{definition}[Maximal sequential subset]
  \label{def:maximal-set-v-s}
  Let $\vertexset{}$ be a set of sub-tasks allocated on a single
  engine, and let $\tau$ be a tagged task such that
  $\vertexset{\tau} \subseteq \vertexset{}$.

  A \emph{maximal sequential subset} $\vertexset{}^{M}$ is a maximal
  subset of $\vertexset{\tau}$ such that none of the sub-tasks in
  $\vertexset{}^{M}$ has a null predecessor. Further, we denote by
  $v^M \in \vertexset{}^{M}$ the sub-task with the shorter local
  deadline in $\vertexset{}^{M}$.
\end{definition}

We observe that, since all the sub-tasks in $\vertexset{}^{M}$ are
allocated on the same engine and since they do not have any
predecessor sub-task allocated on a different engine (no empty
predecessor), they can be activated as soon as the predecessor
sub-tasks have finished.

Now, suppose $v_1, v_2 \in \vertexset{}^{M}$ and that $v_1$ is an
immediate predecessor of $v_2$. If $v_1$ preempts a sub-task $v_j$,
and $D(v_2) \leq D(v_j)$, then $v_j$ can be executed only after $v_2$
has finished. This means that the cost of preempting $v_j$ can be
accounted to only one between $v_1$ and $v_2$. We assign this
preemption cost to the sub-task $v^M$ with the shorter local deadline
among all sub-tasks of $\vertexset{}^{M}$, whereas the others do not
pay any preemption cost.  The preemption cost of any other sub-task in
$\vertexset{}'$ is set equal to $0$. For all sub-tasks that have a
null predecessor, we compute a preemption cost as in Lemma
\ref{theo:worst_preemption}.

Finally, for any tagged task graph $\tau$, the preemption cost of one
of its sub-tasks $v_i \in \vertexset{\tau}$ can be computed as
follows:
\begin{itemize}
\item If $v_i = v^M$, or if $v_i$ has a null predecessor, then 
  \begin{equation}
    \label{eq:preemption_cost_vm}
    \mathsf{pc}^i = \max\{ \mathsf{pc}(v) | v \in \vertexset{} \setminus \vertexset{\tau} \wedge D(v) > D(v_s) \};
  \end{equation}
\item otherwise, 
  \begin{equation}
    \label{eq:preemption_cost_novm}
    \mathsf{pc}^i = 0    
  \end{equation}
\end{itemize}

\begin{theorem}\label{theo:preempt-all}
  Let $\vertexset{}=\{v_{1},v_{2}, \cdots , v_{K}\}$ be a set of
  sub-tasks scheduled by to EDF. Consider
  $\vertexset{}^{\mathsf{pc}}=\{v_{1}',v_{2}', \cdots , v_{K}'\}$
  where $v'_i$ has the same parameters as $v_i$, except for the wcet
  that is computed as $\charge{v'_i} = \charge{v_i} + \mathsf{pc}^i$,
  and $\mathsf{pc}^i$ is computed as in Equation
  \eqref{eq:preemption_cost_vm} or \eqref{eq:preemption_cost_novm}.
  If $\vertexset{}^{\mathsf{pc}}$ is schedulable by EDF when
  considering a null preemption cost, then $\vertexset{}$ is
  schedulable when considering the cost of preemption.
\end{theorem}
\begin{proof}
  We report here a proof sketch. 

  Consider any non-source sub-task $v_i \in \vertexset{}^M$: it is
  activated as soon as the preceding sub-tasks have finished executing
  their corresponding instances. Then, if one of the preceding task of
  $v_i$ preempted a task $v_j$, the preemption cost has already been
  accounted in the worst-case execution time of the preceding task; as
  discussed above $v_j$ can only resume execution after $v_i$ has
  completed. Thus, no further preemption cost need to be accounted.

  If instead none of the preceding sub-task of $v_i$ has preempted
  $v_j$, then $v_j$ cannot start executing before $v_i$ completes
  because its deadline is not smaller than $D(v_i)$, hence no
  preemption will occur.

  In any case, no cost of preemption needs to be accounted for to
  $v_i$.
\end{proof}

%% file: figs/sep_1.tex
\begin{tikzpicture}
  \def\d{1.7}
  \node at (0,0) [circle,draw] (v_1) {\tiny $v_{1}^{\mathsf{CPU}}$};
  \node at (\d,0) [circle, draw] (v_2) {\tiny $v_{2}^{\mathsf{CPU}}$};
  \node at (\d/2,-0.5*\d) [diamond,draw] (F_2) {F};
  \node at (\d/2,-2.5*\d) [circle, draw] (v_10) {\tiny $v_{8}^{\mathsf{CPU}}$};
  
  \node at (\d/2,-1.5*\d) [diamond,draw,fill] (F_3) {F};
  \node at (0,-1*\d) [circle, draw] (v_7) { $\emptyset$};
  \node at (\d,-1*\d) [circle, draw] (v_8) { $\emptyset$};
  \draw [<-](F_2) -- (v_1);
  \draw [<-](F_2) -- (v_2);
  \draw [->](F_2) -- (v_7);
  \draw [->](F_2) -- (v_8);
  \draw [<-](F_3) -- (v_7);
  \draw [<-](F_3) -- (v_8);
   \draw [<-](v_10) -- (F_3);
\end{tikzpicture}


%% file: figs/sep_2.tex
\begin{tikzpicture}
  \def\d{1.7}
  \node at (0,0) [circle,draw] (v_1) { $\emptyset$};
  \node at (\d,0) [circle, draw] (v_2) { $\emptyset$};
  \node at (\d/2,-0.5*\d) [diamond,draw] (F_2) {F};
  \node at (\d/2,-2.5*\d) [circle, draw] (v_10) { $\emptyset$};
  
  \node at (\d/2,-1.5*\d) [diamond,draw,fill] (F_3) {F};
  \node at (0,-1*\d) [circle, draw] (v_7) {\tiny $v_{6}^{\mathsf{DLA}}$};
  \node at (\d,-1*\d) [circle, draw] (v_8) { $\emptyset$};
  \draw [<-](F_2) -- (v_1);
  \draw [<-](F_2) -- (v_2);
  \draw [->](F_2) -- (v_7);
  \draw [->](F_2) -- (v_8);
  \draw [<-](F_3) -- (v_7);
  \draw [<-](F_3) -- (v_8);
   \draw [<-](v_10) -- (F_3);
\end{tikzpicture}


%% file: figs/sep_3.tex
\begin{tikzpicture}
  \def\d{1.7}
  \node at (0,0) [circle,draw] (v_1) { $\emptyset$};
  \node at (\d,0) [circle, draw] (v_2) { $\emptyset$};
  \node at (\d/2,-0.5*\d) [diamond,draw] (F_2) {F};
  \node at (\d/2,-2.5*\d) [circle, draw] (v_10) { $\emptyset$};
  
  \node at (\d/2,-1.5*\d) [diamond,draw,fill] (F_3) {F};
  \node at (0,-1*\d) [circle, draw] (v_7) { $\emptyset$};
  \node at (\d,-1*\d) [circle, draw] (v_8) {\tiny $v_{7}^{\mathsf{dGPU}}$};
  \draw [<-](F_2) -- (v_1);
  \draw [<-](F_2) -- (v_2);
  \draw [->](F_2) -- (v_7);
  \draw [->](F_2) -- (v_8);
  \draw [<-](F_3) -- (v_7);
  \draw [<-](F_3) -- (v_8);
   \draw [<-](v_10) -- (F_3);
\end{tikzpicture}


%% file: figs/task_off_d.tex
\begin{tikzpicture}

\draw [line width=0.5mm](0,0)--(22.5,0);
\draw [line width=0.5mm] (0,-3)--(22.5,-3);

\draw [line width=0.5mm] (0,3)--(22.5,3);

\draw[->] [line width=0.5mm] (1,3) -- (1,4.5) ;
\draw[<->] [line width=0.5mm] (6,3) -- (6,4) ;
\draw[<-] [line width=0.5mm] (16,3) -- (16,4) ;
\draw[<-] [line width=0.5mm] (21,3) -- (21,4.5) ;

\draw[->] [line width=0.5mm] (6,0) -- (6,1) ;
\draw[<-] [line width=0.5mm] (16,0) -- (16,1) ;

\draw[->] [line width=0.5mm] (6,-3) -- (6,-2) ;
\draw[<-] [line width=0.5mm] (16,-3) -- (16,-2) ;

\filldraw[fill=red, draw=black,opacity=0.3] (16,3) rectangle(18.5,3.5);
\draw(16,3) rectangle(18.5,3.5);

\filldraw[fill=red, draw=black,opacity=0.3] (1.5,3) rectangle(3.5,3.5);
\filldraw[fill=red, draw=black,opacity=0.3] (3.5,3) rectangle(6,3.5);


\draw (1.5,3) rectangle(3.5,3.5);
\draw (3.5,3) rectangle(6,3.5);

\filldraw[fill=blue, draw=black,opacity=0.3,dashed] (6,0) rectangle(11,0.5);
\draw[dashed] (6,0) rectangle(11,0.5);

\filldraw[fill=green, draw=black,opacity=0.6,dashed] (6,-3) rectangle(8.5,-2.5);
\draw[dashed] (6,-3) rectangle(8.5,-2.5);

\node  at (17.25,2.5) {\huge $v_8$};
\node  at (2.25,2.5) {\huge $v_1$};
\node  at (4.5,2.5) {\huge $v_2$};
\node  at (-1,3) {\Large \sf CPU};
\node  at (-1,0) {\Large \sf iGPU};
\node  at (-1,-3) {\Large \sf dGPU};
\node  at (8,-0.5) {\huge $v_6$};
\node  at (7.4,-3.5) {\huge $v_7$};

\draw[<->,dashed, line width= 0.5mm] (1,-4.5) -- (16,-4.5);
\draw[dotted] (1,-5.5) -- (1,5.5);
\draw[fill, white] (5,-4) rectangle (9.7,-5.5);
\node (ld) at (7.4,-4.5) {\Large \sf $v_7$ Local deadline};

\draw[<->,dashed, line width= 0.5mm] (6,-2.1) -- (16,-2.1);
\draw[fill, white] (8.4,-2) rectangle (13.4,-2.2);
\node (ld) at (10.8,-2.1) {\Large \sf $v_7$ relative deadline};
\node (ld) at (6,1.3) {\Large \sf O$(v_6)$};

\node (ld) at (21,5) {\Large \sf Absolute deadline};

\node (ld) at (1,6) {\Large \sf Activation time};

\draw[<->,dashed, line width= 0.5mm] (1,4.6) -- (21,4.6);
\draw[fill, white] (8.,4.2) rectangle (13.4,5);
\node (ld) at (10.8,4.6) {\Large \sf task  relative deadline};

\end{tikzpicture}

%% file: texfiles/allocation.tex
\section{Allocation}
\label{sec:allocation}


\subsection{Allocation of task specifications}
\label{sec:vertex-allocation}

The goal of our methodology is to allocate a set of task
specifications into a set of engines, by selecting alternative
implementations, so that all tasks complete before their
deadlines. From a operational point of view, is is equivalent to
finding a feasible solution to a complex Integer Linear Programming
problem. In facts, given the large number of combinations (due to
alternative nodes, condition-control nodes, and allocation decisions),
an ILP formulation of this problem fails to produce feasible solutions
in an acceptable short time. Therefore, in this section we propose a
set of greedy heuristics to quickly explore the space of solutions.

Algorithm \ref{algo:alloc-task-spec} describes the basic methodology
of our approach. The algorithm can be customised with four parameters:
$\mathsf{oder}$ is the sorting order of the concrete task sets (see
Sections \ref{sec:graph-spec-transf} and \ref{sec:graph-separation});
parameter $\mathsf{slack}$ concerns the way the slack is distributed
when assigning intermediate deadlines and offsets (see Section
\ref{sec:artif-deadl-offs}); parameter $\mathsf{alloc}$ can be
best-fit (BF) or worst-fit (WF); parameter $\mathsf{omit}$ concerns
the strategy to eliminate sub-tasks when possible (see Section
\ref{sec:parallel-allocation}).

At each step, the algorithm tries to allocate one single task
specification (for loop at line 3). For each task, it first generates
all concrete tasks (line 4), and sorts them according to one
relationship order ($\succ$ or $\gg$). Then, for each concrete task,
if first assigns the intermediate deadlines and offsets according to
the methodology described in Section \ref{sec:artif-deadl-offs} (line
9), using one between the fair or the proportional slack
distributions. Then, it separates the concrete tasks into tagged tasks
according to the corresponding tags (line 10). 

Then, the algorithm tries to allocate every tagged task onto single
engines having the corresponding tag (line 14) (this procedure is
described below in Algorithm \ref{algo:alloc-seq-spec}). If a feasible
allocation is found, the allocation is generated, and the algorithm
goes to the next specification task (lines 15-16). If no feasible
sequential allocation can be found, the next concrete task is tested.

\begin{algorithm}[h]
  \caption{Allocation algorithm} \label{algo:alloc-task-spec}
  \begin{algorithmic}[1]
        \State \textbf{input :} $\taskset{}$: set of task specifications
        \State \textbf{parameters :} $\mathsf{order}$ ($\succ$ or $\gg$), $\mathsf{slack}$ (fair or proportional),  
                     \\\hspace{2.07cm}$\mathsf{alloc}$ (BF or WF), $\mathsf{omit}$ (parallel or random) 
        \State \textbf{output :} SUCCESS or FAIL
        \For {$\task{} \in \taskset{}$}             
             \State $\Omega$ = generate\_concrete\_task($\task{}$)
             \State $\mathsf{sort}(\Omega, \mathsf{order})$ 
             \For {$(\concrete{\task{}} \in \Omega)$}
                  \State assign\_deadlines\_offsets($\concrete{\task{}}$, $\mathsf{slack}$)
                  \State $\mathcal{S}(\concrete{\task{}})$ = filter\_tagged\_task($\concrete{\task{}}$) 
             \EndFor
             \State allocated = \textbf{false}
             \For {$(\concrete{\task{}} \in \Omega)$}
                  \If {(feasible\_sequential($\mathcal{S}(\concrete{\task{}})$, $\mathsf{alloc}$))}
                      \State allocated = \textbf{true}; assign sub-tasks to engines
                      \State \textbf{break};
                  \EndIf
              \EndFor
              \If {(\textbf{not} allocated)}
                 \For {($\concrete{\task{}} \in \Omega$)}
                     \State $(\concrete{\task{}}',\concrete{\task{}}'')$ = parallelize($\concrete{\task{}}$, $\mathsf{alloc}$, $\mathsf{omit}$)
                      \If {($\concrete{\task{}}' \neq \emptyset$)}
                          \State allocate $\concrete{\task{}}'$ to selected engines
                          \State add back $\concrete{\task{}}''$ to $\taskset{}$
                          \State allocated = \textbf{true}
                          \State \textbf{break}
                      \EndIf
                \EndFor
                \State \textbf{if} (\textbf{not} allocated) \textbf{then} \textbf{return} FAIL
              \EndIf
         \EndFor
         \State \textbf{return} SUCCESS
  \end{algorithmic}
\end{algorithm}

The algorithm gives priority to single-engine allocations because they
reduce preemption cost, as discussed in
Section~\ref{sec:preempt-aware-sched}. In particular, by allocating an
entire tagged task onto a single engine, we reduce the number of null
sub-task to the minum necessary, and so we can assign the cost of
preemption to fewer sub-tasks.

If none of the concrete tasks of a specification task can be allocated
(line 17), this means that one of the tagged tasks could not be
allocated on a single engine. Therefore, the algorithms tries to break
some of the tagged tasks of a concrete task into parallel tasks to be
executed on different engines of the same type. This is performed by
procedure \textsf{parallelize}, which will be described in Section
\ref{sec:parallel-allocation}. In particular, one part of the concrete
task will be allocated, while the second part will be put back in the
list of not-yet-allocated task graphs (line 24).

If also this process is unable to find a feasible concrete task, the
analysis fails (line 29).

\newcommand{\Input}[1]{\State {\bf input:}  #1}
\newcommand{\Output}[1]{\State {\bf output:} #1}

\subsection{Sequential allocation}
\label{sec:sequ-alloc}

Algorithm \ref{algo:alloc-seq-spec} tries to allocate a concrete task
on a minimal number of engines. It takes as input a set of tagged
tasks. For each tagged task, it selects the corresponding engines, and
sorts them according to the $\mathsf{alloc}$ parameter, that is in
decreasing order of utilization in the case of Best-Fit, or in
increasing order of utilization in case of Worst-Fit. Then, it tests
the feasibility of allocating the tagged task on each engine in
turn. If the allocation is successful, the next tagged task is tested,
otherwise the algorithm tries the next engine. If the tagged task
cannot be allocated on any engine, the algorithm fails. If all tagged
tasks have been allocated, the corresponding allocation is returned.


\begin{algorithm}[h]
\caption{feasible\_sequential}\label{algo:alloc-seq-spec}
\begin{algorithmic}[1]
  \Input{$\mathcal{S}(\concrete{\task{}})$: set of tagged tasks, $\mathsf{alloc}$}
  \Output{feasibility: \textbf{SUCCESS} or \textbf{FAIL}}
  \For {$(\concrete{\task{}}({\mathsf{tag}}) \in \mathcal{S}(\concrete{\task{}}))$}
        \State engine\_list=select\_engine($\mathsf{tag}$) 
        \State sort\_engines(engine\_list, $\mathsf{alloc}$) 
        \State $f$ = \textbf{false}
        \State nfeas = 0
        \For {$(e \in  \mathsf{engine\_list})$}
               \State $f$ = dbf\_test($\concrete{\task{}} \cup \taskset_e$)
               \If {($f$)}
                   \State save\_allocation($\concrete{\task{}}$, $e$)
                   \State nfeas ++
                   \State {\bf break}
               \EndIf
        \EndFor
        \State \textsf{if} (\textbf{not} $f$) \textbf{then} \textbf{return} FAIL;     
   \EndFor  
   \If {(nfeas $= |\mathcal{S}(\concrete{\task{}})|$)}
         \State \textbf{return} SUCCESS, saved\_allocations
   \EndIf
  \end{algorithmic}
\end{algorithm}

\subsection{Parallel allocation}
\label{sec:parallel-allocation}

When the sequential allocation fails for a given task specification,
the algorithm tries to allocate one or more of its tagged tasks onto
multiple engines having the same tag. Algorithm
\ref{algo:paralleliz-concrete} takes as input a concrete task and two
parameters, $\mathsf{alloc}$ for BF or WF heuristics, and
$\mathsf{omit}$ to select which sub-task to remove first. 

For each tagged task of the concrete task (line 5), the algorithm
selects the list of engines corresponding to the selected tag, and
sorts them according to BF or WF (line 7). Then, it tries to test the
feasibility of the tagged task on each engine (line 9). If the test
fails, it removes one sub-task from the tagged task and adds it to
list of non allocated sub-tasks $\concrete{\tau}''$ (line 11). We
propose two heuristics: 
\begin{enumerate}
\item \textbf{Random} heuristic: it selects a random sub-task and adds
  it to the omitted list.
\item \textbf{Parallel} heuristic: to be feasible, the critical path
  of each tagged task must be feasible even on a unlimited number of
  engines. Thus, we are interested in sub-tasks that do not belong
  to the critical path because they are the ones causing the
  non-feasibility. Thus, they are omitted one by one until finding a
  feasible schedule.
\end{enumerate}
The feasibility test is repeated until a feasible subset of
$\concrete{\tau}(\mathsf{tag})$ is found. The omitted tasks are tried
on the next engine with the same tag (line 16). At the end of the
procedure, two concrete tasks are produced, $\concrete{\tau}'$ is the
feasible part that will be allocated, while $\concrete{\tau}''$ will
be tried again in the following iteration of Algorithm
\ref{algo:alloc-task-spec}. 



\begin{algorithm}[h]
\caption{parallelize}\label{algo:paralleliz-concrete}
\begin{algorithmic}[1]
  \Input{$\concrete{\task{}}$: concrete task, $\mathsf{alloc}$ (BF or WF),}
         \\\hspace{0.95cm} $\mathsf{omit}$ (parallel or random) 
  \Output{concrete tasks $(\concrete{\tau}', \concrete{\tau}'')$}

  \State{$\concrete{\tau}' = \emptyset, \concrete{\tau}''=\emptyset$}
  \For {$(\concrete{\task{}}(\mathsf{tag}) \in  \mathcal{S}(\concrete{\task{}}))$}
      \State engine\_list=select\_engines(tag) 
      \State sort(engine\_list, $\mathsf{alloc}$) 
      \For {$(e \in  \mathsf{engine\_list})$} 
          \State $f$=dbf\_test($\concrete{\task{}}(\mathsf{tag}) \cup \taskset_{e}$)
          \While {(\textbf{not} $f$) }
             \State $\concrete{\tau}''$ = $\concrete{\tau}'' \cup$ remove($\concrete{\task{}}(\mathsf{tag}),\mathsf{omit}$)
             \State $f$=dbf\_test($\concrete{\task{}}(\mathsf{tag}) \cup \taskset_{\mathsf{E}}$)
          \EndWhile
          \If {($\concrete{\task{}}(\mathsf{tag}) \neq \emptyset$ )}
             \State $\concrete{\tau}' = \concrete{\tau}' \cup$ save\_allocation($\concrete{\task{}}(\mathsf{tag}),e$)
             \State $\concrete{\tau}(\mathsf{tag}) = \concrete{\tau}'', \concrete{\tau}'' = \emptyset$
             \State allocated = \textbf{true}
             \State \textbf{break}
          \EndIf
      \EndFor
      \State \textbf{if} (\textbf{not} allocated) \textbf{return} $\emptyset, \concrete{\tau}$
 \EndFor
 \State \textbf{return} $\concrete{\tau}', \concrete{\tau}''$
\end{algorithmic}
\end{algorithm}





%% file: texfiles/stat.tex
\section{Related work}
\label{sec:related-work}

Many authors
\cite{qamhieh2013global,melani_bertogna,saifullah_multi-core_2013,li2014analysis,saifullah2012real,fonseca2016response,Marinca:2004:ADA:2285761.2285882,peppehoussemJSA,houssamPeppeRTNS}
have proposed real-time task models based on DAGs. 
However, to the best of our knowledge, none of
the existing models supports alternative implementations of the same
functionality on different computing engines.

Authors of \cite{Marinca:2004:ADA:2285761.2285882} studied the
deadline assignment problem in distributed real-time systems. They
formalize the problem and identify the cases where deadline assignment
methods have a strong impact on system performances. They propose Fair
Laxity Distribution (FLD) and Unfair Laxity Distribution (ULD) and
study their impact on the schedulability. In \cite{li2014analysis},
authors analyze the schedulability of a set of DAGs using global EDF,
global rate-monotonic (RM), and federated scheduling. In
\cite{wu2014deadline}, the authors present a general framework of
partitioning real-time tasks onto multiple cores using resource
reservations. They propose techniques to set activation time and
deadlines of each task, and they an use ILP formulation to solve the
allocation and assignment problems. However, when applying such
approaches on large applications consisting of hundred of sub-tasks,
the analysis can be highly time consuming. 

DAG fixed-priority partitioned scheduling has been presented in
\cite{fonseca2016response}. The authors propose methods to compute a
response time with tight bounds. They present partitioned DAGs as a
set of self-suspending tasks, and proposed an algorithm to traverse a
DAG and characterize the worst-case scheduling scenario.

Unlike previous models, Melani et al \cite{melani_bertogna} proposed
to model conditional branches in the code in a way similar to our
conditional nodes, however their model is not able to express off-line
alternative patterns. They proposed different methods to compute an
upper-bound on the response-time under global scheduling algorithms.
In \cite{stigge2011digraph}, alternative on-line execution patterns
can be expressed using \emph{digraph}s. However, the digraph model
cannot express parallelism and only supports sequential tasks.

In this paper we assume preemptive EDF scheduling. Typically,
preemption on classical CPUs can be assumed to be a negligible
percentage of the task execution. However, this is not always the case
with GPUs processors. Depending on the computing architecture and on
the nature of the workload, GPU tasks present different degrees of
preemption granularity and related preemption costs. Initial work on
preemptive scheduling on GPUs assumed preemption was viable at
the \emph{kernel} granularity \cite{zhong2014kernelet}. A finer
granularity for computing workloads is represented by CTA (Cooperative
Thread Array) level preemption, hence, preemption occurs at the
boundaries of group of parallel threads that execute within the same
GPU computing cluster \cite{amert2017gpu,capodieci2017sigamma}. 
In such a scenario, the cost of preempting an executing context on a
GPU might present significant differences as it will involve saving
and restoring contexts of variable size and/or reaching the next
viable preemption point. Overhead measurements operated in the cited
contributions calls for modeling each GPU sub-task with a specific
non-negligible preemption cost that can be in the same order of
magnitude of the execution time of the sub-task.



%% file: texfiles/results.tex
\section{Results and discussions}
\label{sec:results}

In this section, we evaluate the performance of our scheduling
analysis and allocation strategies. We compare against the model
cp-DAG proposed by Melani et al. \cite{melani_bertogna}. Please notice
that in \cite{melani_bertogna} the authors proposed an analysis for
cp-DAGs in the context of global scheduling, whereas our analysis is
based on partitioned scheduling. Therefore, we extended the cp-DAG
model to support multiple engines by adding a randomly selected tag to
each node of the graph. Moreover we applied the same allocation
heuristics of Section \ref{sec:allocation} and the same scheduling
analysis of Section \ref{sec:analysis} to C-DAGs and to cp-DAG. 


In the following experiments, we considered the NVIDIA Jetson AGX
Xavier\footnote{ \url{https://elinux.org/Jetson_AGX_Xavier}}. It
features 8 CPU cores, and four different kinds of accelerators: one
discrete and one integrated GPU, one DLA and one PVA. Each accelerator
is treated as a single computing resource. In this way, we can exploit
task level parallelism as opposed to allowing the parallel execution
of more than one sub-task to partitions of the accelerators (e.g: at a
given time instant, only one sub-task is allowed to execute in all the
computing clusters of a GPU).

\subsection{Task Generation}
\label{sec:task-generation}

We apply our heuristics on a large number of randomly generated
synthetic task sets. 

The task set generation process takes as input an engine/tag
utilization for each tag on the platform. First, we start by
generating the utilization of the $n$ tasks by using the
UUniFast-Discard \cite{uni} algorithm for each input
utilization. Graph sub-tasks can be executed in parallel, thus task
utilization can be greater than 1. The sum of every per-tag
utilization is a fixed number upper bounded by the number of engines
per tag.

The number of nodes of every task is chosen randomly between 10 and
30. We define a probability $p$ that expresses the chance to have an
edge between two nodes, and we generate the edges according to this
probability. We ensure that the graph depth is bounded by an integer
$d$ proportional to the number of sub-tasks in the task. We also
ensure that the graph is \emph{weakly connected} (i.e. the
corresponding undirected graph is connected); if necessary, we add
edges between non-connected portions of the graph. Given a sub-task
node, one of its successors is an alternative node or a conditional
node with probability of $0.7$.



To avoid untractable hyper-periods, the period of every task is
generated randomly according from the list, where the minimum is $120$
and the maximum is $120,000$. For every sub-task, we randomly select a
tag. Further, for each tag, we use algorithm UUNIFAST discard again to
generate single sub-task utilization. Thus, the sub-task utilization
can never exceed 1. Further, we inflate the utilization of each
sub-task by the task period to generate the worst case execution time
of every vertex.

A cp-DAG is generated from a C-DAG by selecting one of the possible
concrete tasks at random.

\subsection{Simulation results and discussions}

We varied the baseline utilization from $0$ to the number of engines
per engine tag in $16$ steps. Therefore, the step size vary from one
engine tag to the other: the step size is $0.5$ for CPUs, and $0.0625$
for the others. For each utilization, we generated a random number of
tasks between 20 and 25.

\begin{figure}[h]
  \centering
  \begin{tikzpicture}[scale=0.8]
    \begin{axis}
      [ axis x line=bottom, axis y line = left, xlabel={Total
        Utilization index}, ylabel={Average scarce-engine utilization},
      legend entries={BRF-P,BOF-P, WRF-P,WOF-P,  BOP-P,BRP-P,WOP-P,WRP-P}, legend
      style={at={(1.35,0.7)}}, ]
      \addplot [mark=diamond*,black]table[y index=1,x index=0]{res/avg_u_sc_0.6_par.dat};
      \addplot [mark=square*,blue]table[y index=2,x index=0]{res/avg_u_sc_0.6_par.dat};
      \addplot [mark=diamond,black] table[y index=3,x index=0]{res/avg_u_sc_0.6_par.dat};
      \addplot [mark=square,blue] table[y index=4,x index=0]{res/avg_u_sc_0.6_par.dat};
      \addplot [mark=o,yellow] table[y index=1,x index=0]{res/ml_a_u_s.dat};
      
    \end{axis}
  \end{tikzpicture}
  \caption{DLA, GPUs, PVA utilizations  vs  total utilization.}\label{fig:sc_a_u}
\end{figure}
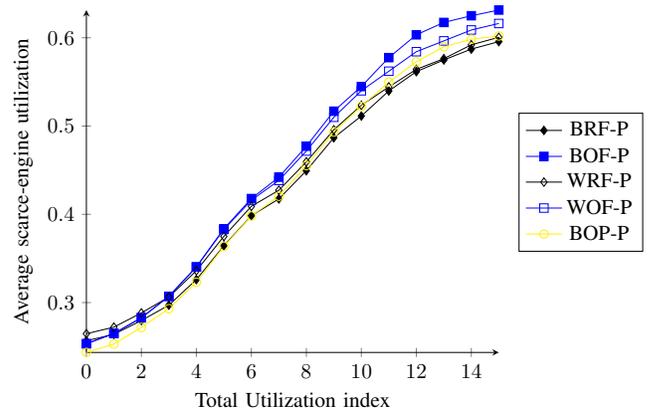

The results are presented as follows. Each algorithm is described
using $3$ letters: (i) the first letter is either $\mathsf{B}$ for
best fit or $\mathsf{W}$ for worst first allocation techniques; (ii)
the second is either $O$ for the $\succ$ order relation, or $R$ for
the $\gg$ order relation; (iii) the third character describes the
deadline assignment heuristic, $F$ for fair and $P$ for
proportional. The algorithm name may also contain either option $P$
for the parallel allocation heuristic that eliminates parallel nodes
first, or $R$ the random heuristic which randomly selects the sub-task
to remove. For
Figures~\ref{fig:sched-rate},~\ref{fig:active_cores},~\ref{fig:acti-u-c},~\ref{fig:sc_a_u},
we run $85$ simulations per utilization step.

\definecolor{cl1}{HTML}{E41A1C}
\definecolor{cl2}{HTML}{377EB8}
\definecolor{cl3}{HTML}{4DAF4A}
\definecolor{cl4}{HTML}{984EA3}

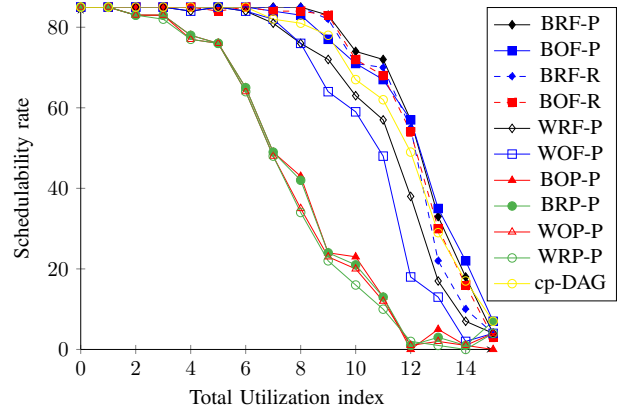
\begin{figure}[h]
  \centering
  \begin{tikzpicture}[scale=0.8]
    \begin{axis}
      [axis x line=bottom, axis y line = left, xlabel={Total
        Utilization index}, ylabel={Schedulability rate},
      legend entries={BRF-P,BOF-P, BRF-R,BOF-R,WRF-P,WOF-P,  BOP-P,BRP-P,WOP-P,WRP-P,cp-DAG}, legend
      style={at={(1.3,1)}}, ]
      \addplot [mark=diamond*,black]table[y index=4,x index=0]{res/sched_rate_0.6_par.dat};
      \addplot [mark=square*,blue]table[y index=2,x index=0]{res/sched_rate_0.6_par.dat};
      \addplot [mark=diamond*,blue,dashed]table[y index=10,x index=0]{res/sched_rate_0.6_par.dat};
      \addplot [mark=square*,red,dashed] table[y index=9,x index=0]{res/sched_rate_0.6_par.dat};
      \addplot [mark=diamond,black] table[y index=8,x index=0]{res/sched_rate_0.6_par.dat};
      \addplot [mark=square,blue] table[y index=6,x index=0]{res/sched_rate_0.6_par.dat};
      \addplot [mark=triangle*,red] table[y index=1,x index=0]{res/sched_rate_0.6_par.dat};
      \addplot [mark=*,cl3]table[,y index=3,x index=0]{res/sched_rate_0.6_par.dat};
      \addplot [mark=triangle,red] table[y index=5,x index=0]{res/sched_rate_0.6_par.dat};
      \addplot [mark=o,cl3] table[y index=7,x index=0]{res/sched_rate_0.6_par.dat};
      \addplot [mark=o,yellow] table[y index=1,x index=0]{res/ml_r.dat};
    \end{axis}
  \end{tikzpicture}
  \caption{Schedulability rate VS total utilization.}\label{fig:sched-rate}
\end{figure}

Figure \ref{fig:sched-rate} represents the schedulability rate of each
combination of heuristics cited above as a function of the total
utilization. The fair deadline assignment technique presents better
schedulability rates compared to proportional deadline assignment. In
general, BF heuristic combinations outperform WF heuristic: this can
be explained by observing that BF tries to pack the maximum number of
sub-tasks into the minimum number of engines, and this allows for more
flexibility to schedule \emph{heavy} tasks on other engines.

In the figures, the cp-DAG model proposed in \cite{melani_bertogna} is
shown in yellow. Since the cp-DAG has no alternative implementations,
the algorithm has less flexibility in allocating the sub-tasks,
therefore \emph{by construction} the results for C-DAG dominate the
corresponding results for cp-DAG. However, it is interesting to
measure the difference between the two models: for example in Figure
\ref{fig:sched-rate} the difference in the schedulability rate between
the two models is between 10\% and 20\% for utilization rates between
6 and 14.

When the system load is low, all combinations of heuristics allow
having high schedulability rates. BRF shows better results because it
is aimed at relaxing the utilization of scarce engines, thus avoiding
the unfeasibility of certain task sets due to a high load on a scarce
engines (DLA and PVA/ GPUs). However, when dealing with a highly
loaded system, BOF presents better schedulability rates, as it reduces
the execution overheads on all engines.

\begin{figure}[h]
  \centering
  \begin{tikzpicture}[scale=0.8]
    \begin{axis}
      [ axis x line=bottom, axis y line = left, xlabel={Total
        Utilization index}, ylabel={\# active CPUs},
      legend entries={BRF-P,BOF-P, WRF-P,WOF-P,  BOP-P,BRP-P,WOP-P,WRP-P}, legend
      style={at={(1.35,0.7)}}, ]
      \addplot [mark=diamond*,black]table[y index=4,x index=0]{res/avg_ncore_0.6_par.dat};
      \addplot [mark=square*,blue]table[y index=2,x index=0]{res/avg_ncore_0.6_par.dat};
      \addplot [mark=diamond,black] table[y index=8,x index=0]{res/avg_ncore_0.6_par.dat};
      \addplot [mark=square,blue] table[y index=6,x index=0]{res/avg_ncore_0.6_par.dat};
      \addplot [mark=triangle*,red] table[y index=1,x index=0]{res/avg_ncore_0.6_par.dat};
      \addplot [mark=*,cl3]table[,y index=3,x index=0]{res/avg_ncore_0.6_par.dat};
      \addplot [mark=triangle,red] table[y index=5,x index=0]{res/avg_ncore_0.6_par.dat};
      \addplot [mark=o,cl3] table[y index=7,x index=0]{res/avg_ncore_0.6_par.dat};
      \addplot [mark=o,yellow] table[y index=1,x index=0]{res/ml_a_c.dat};
    \end{axis}
  \end{tikzpicture}
  \caption{\#Active CPUS vs total utilization.}\label{fig:active_cores}
\end{figure}
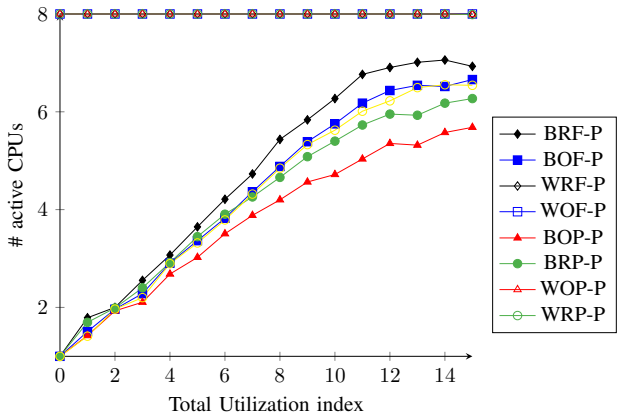

Figure \ref{fig:active_cores} reports the average number of active
cores (CPUs) as a function of the total utilization. WF-based
heuristics always use the highest number of CPU cores because our task
generator outputs at least $15$ CPU subtasks. Hence, the number of
tasks is larger than the available number of CPU cores (which is 8, in
our test platform). BF heuristics allows to pack the maximum number of
sub-tasks on the minimum number of engines, thus the utilization
increases quasi-linearly. This occurs until the maximum schedulability
limit is reached (i.e. number of cores). BRF heuristic uses more CPU
cores because it \emph{preserves} the scarce resources, thus it uses
more CPU engines. As BOF privileges reducing the overall load, it
reduces the load on the CPUs compared to BRF.

\begin{figure}[h]
  \centering
  \begin{tikzpicture}[scale=0.8]
    \begin{axis}
      [ axis x line=bottom, axis y line = left, xlabel={Total
        Utilization index}, ylabel={active CPUs utilization},
      legend entries={BRF-P,BOF-P, WRF-P,WOF-P,  BOP-P,BRP-P,WOP-P,WRP-P}, legend
      style={at={(1.35,0.7)}}, ]

      \addplot [mark=diamond*,black]table[y index=4,x index=0]{res/avg_u_a_0.6_par.dat};
      \addplot [mark=square*,blue]table[y index=2,x index=0]{res/avg_u_a_0.6_par.dat};
      \addplot [mark=diamond,black] table[y index=8,x index=0]{res/avg_u_a_0.6_par.dat};
      \addplot [mark=square,blue] table[y index=6,x index=0]{res/avg_u_a_0.6_par.dat};
      \addplot [mark=triangle*,red] table[y index=1,x index=0]{res/avg_u_a_0.6_par.dat};
      \addplot [mark=*,cl3]table[,y index=3,x index=0]{res/avg_u_a_0.6_par.dat};
      \addplot [mark=triangle,red] table[y index=5,x index=0]{res/avg_u_a_0.6_par.dat};
      \addplot [mark=o,cl3] table[y index=7,x index=0]{res/avg_u_a_0.6_par.dat};
      \addplot [mark=o,yellow] table[y index=1,x index=0]{res/ml_a_u.dat};
    \end{axis}
  \end{tikzpicture}
  \caption{Active CPU utilization VS total utilization}\label{fig:acti-u-c}
\end{figure}
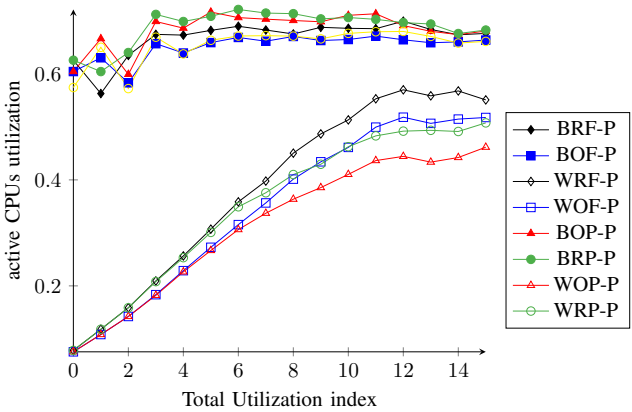

Figure \ref{fig:acti-u-c} shows the average active utilization for
CPUs. Average utilization of BF-based heuristics is higher compared to
WF. In fact, the latter distributes the work on different engines thus
the per-core utilization is low in contrast to BF. Again, BRF has
higher utilization than BOF because it schedules more workload on CPU
cores than the other heuristics. As the workload is equally
distributed on different CPUs, the WF heuristics may be used to reduce
the CPUs operating frequency to save dynamic energy. Regarding BF
heuristics, we see that BRF is not on the top of the average load
because it uses more cores than the others.

Figure \ref{fig:sc_a_u} shows the average utilization of the scarce
resources. As you may notice, order relation $\gg$ based heuristics
allows to reduce the load on the scarce resources compared to
$\succ$. In fact, the higher is the load, the less loaded are the
scarce resources.

\subsection{Preemption cost simulation}
\label{sec:preempt-cost-simul}



In all previous experiments, we applied the analysis described in
Section~\ref{sec:preempt-aware-sched} to account for preemption
costs. In particular, we applied the technique of
Theorem~\ref{theo:preempt-all}, by assuming that the cost of
preempting a sub-task is 30\% of the sub-task execution time on a GPU,
10\% on DLA and PVA, and 0.02\% on the CPUs. DLA and PVA are
non-preemptable engines, however longer jobs might be split into
smaller chunks and this translates in a splitting overhead as we
submit many kernel calls as opposed of a single batch of commands.
 


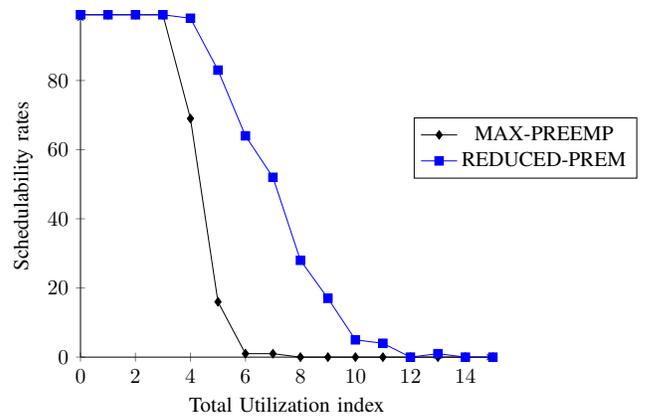
\begin{figure}[h]
  \centering
  \begin{tikzpicture}[scale=0.8]
    \begin{axis}
      [ axis x line=bottom, axis y line = left, xlabel={Total
        Utilization index}, ylabel={Schedulability rates},
      legend entries={MAX-PREEMP,REDUCED-PREM}, legend
      style={at={(1.35,0.7)}}, ]
      \addplot [mark=diamond*,black]table[y index=1,x index=0]{res/preemp.dat};
      \addplot [mark=square*,blue]table[y index=2,x index=0]{res/preemp.dat};
      
    \end{axis}
  \end{tikzpicture}
  \caption{Preemption cost Theorem vs max}\label{fig:preemp_cost}
\end{figure}

To highlight the importance of a proper analysis of the cost of
preemption, in Figure \ref{fig:preemp_cost} we report the
schedulability rates obtained by BRF-P in two different cases: when
considering the analysis of Lemma~\ref{theo:worst_preemption} (where
the maximum preemption cost is charged to all preempting sub-tasks)
and that of Theorem \ref{theo:preempt-all}, where the cost is only
charged to one of the sub-tasks in the maximal sequential subset.


With the increase of of utilization, schedulability drastically falls
for the first method, while the improved analysis of Theorem
\ref{theo:preempt-all} keeps high schedulability rates.


%% file: texfiles/conclu.tex
\section{Conclusions and future work}
\label{sec:concl-future-work}

In this paper, we presented the C-DAG real-time task model, which
allows to specify both off-line and on-line alternatives, to fully
exploit the heterogeneity of complex embedded platforms. We also
presented a scheduling analysis and a set of heuristics to allocate
C-DAGs on heterogeneous computing platforms. The analysis takes into
account the cost of preemption that may be non-negligible in certain
specialized engines.

Results of our extensive synthetic simulations show that a significant
reduction in pessimism occurs with our proposed approach. This lead to
an increase in resource utilization compared to similar approaches in
the literature. As for future work, we are considering extending our
framework to account for memory interference between the different
compute engines, as it is known to cause significant variations in
execution times~\cite{ali2017protecting, cavicchioli2017memory}.